\documentclass[aps,pra,10pt,column,floatfix,showpacs,numerical,footinbib,longbibliography]{revtex4-1}
\usepackage{float}       
\usepackage{amsmath}
\usepackage{amssymb}
\usepackage{dsfont}
\usepackage{bbold} 
\usepackage[%
colorlinks=true,
urlcolor=blue,
linkcolor=blue,
citecolor=blue]{hyperref}
\usepackage{tikz}
\usepackage{graphicx}
\usepackage{subfigure}
\usepackage{svg}
\usepackage{mathtools}
\usepackage{relsize}

\begin{document}
	\renewcommand{\thefootnote}{\fnsymbol {footnote}}
	
	\title{\textbf{Spin-outcome correlations and memory-assisted entropic uncertainty\\in Bhabha scattering with an entangled spectator}}
	\author{Zan Cao}
	\author{Meng-Long Song}
	\author{Xue-Ke Song}
	\author{Liu Ye}
	\author{Dong Wang}
	\email{dwang@ahu.edu.cn}
	\affiliation{State Key Laboratory of Opto-Electronic Information Acquisition and Protection Technology, School of Physics, \textcolor{blue}{Anhui University}, Hefei 230601,  China}

	\begin{abstract}	 
	We investigate spin-outcome correlations and memory-assisted entropic uncertainty in tree-level Bhabha scattering, where an incident electron $A$ scatters with a positron $B$ initially entangled with a spectator electron $C$. 
	 For the fixed-angle post-selected final state, we compare the total memory-assisted uncertainty $U_A$ for three alternative Pauli measurements on $A$ with the entanglement entropy $S(\rho_A)$ across the $A|BC$ partition. 
	 We resolve $U_A$ into the separate uncertainties $U_B$ and $U_C$ and evaluate the corresponding outcome-memory mutual-information sums $G_B$ and $G_C$. This separates changes in the randomness of the outcomes on $A$ from changes in their correlations with each memory. In the parameter regions studied, enhanced $A|BC$ entanglement is accompanied by lower $U_A$, most prominently near high-energy backscattering. 
     An analytic treatment of this limit shows that $B$ factorizes from $AC$: $G_B$ vanishes, whereas $G_C$ increases with the initial $B$-$C$ entanglement. 
	 An entropic complete complementarity analysis characterizes the accompanying loss of local helicity predictability in $\rho_A$ and changes in the coherence, predictability, and mixedness of the interacting pair $\rho_{AB}$.
	 These results show that initial $B$-$C$ entanglement can lower the total memory-assisted uncertainty in high-energy backscattering by strengthening spin-outcome correlations with the spectator $C$, thereby identifying the distinct roles of the interacting particle and the spectator as quantum memories.

	\end{abstract}
	\maketitle
	
	\section{INTRODUCTION}\label{I}
	
	The study of quantum correlations in relativistic particle-physics settings provides a natural bridge between high-energy physics and quantum information theory
	\cite{1.1,1.2,1.3,1.4,1.5,1.6,1.7,1.8ATLAS}.
	In elementary-particle reactions, the helicity states of the outgoing particles can become correlated through the helicity structure of the scattering amplitudes, so that the correlation structure of the final state reflects the underlying interaction dynamics
	\cite{1.9,1.10,5.0SciPostPhys.3.5.036}.
	This perspective has motivated extensive studies of entanglement, coherence, Bell nonlocality, quantum discord,  steering, entropic uncertainty, and quantum Fisher information in a variety of particle-physics settings.
	Representative examples include neutrino oscillations
	\cite{2.1,2.2,2.3,2.4,2.5,2.6,2.7,2.8,2.9,2.10,2.11,2.12},
	meson systems
	\cite{meson3.1,meson3.1.1,meson3.2,meson3.3,meson3.4},
	heavy-quark production at colliders
	\cite{2021PhysRevLett.127.161801,Afik2021,3.5Afik2022quantuminformation,3.6,3.7fhkc-kfhr,5.1,5.2EUR,5.3},
	and other high-energy scattering processes
	\cite{scattering4.2,scattering4.4}.
	
	Among elementary QED processes, Bhabha scattering provides a particularly tractable setting for investigating the generation of helicity correlations. Its tree-level helicity amplitudes depend on the scattering kinematics, particularly the center-of-mass energy and scattering angle, thereby shaping the helicity-correlation structure of the outgoing electron-positron pair
	\cite{5.0SciPostPhys.3.5.036,5.0Cervera2019}.
	Building on this helicity-based framework, recent studies have investigated entanglement generation and saturation, as well as complete complementarity relations, in tree-level QED processes
	\cite{1.10,2.11Entanglementsaturation2025,5.1TreelevelQRDPhysRevD.111.016007}.
	
	Entangled spectators provide a setting for studying how QED scattering reshapes correlations beyond the directly interacting pair. Earlier studies considered such spectators in the inelastic process $e^-e^+\to\mu^-\mu^+$, examining changes in the spectator's local spin state and the redistribution of correlations among the particles \cite{6.1spectatorPhysRevD.2019,6.2spectatorPhysRevD.2022}. In the Bhabha configuration considered here, an electron $A$ scatters with a positron $B$ that is initially entangled with a spectator electron $C$. The redistribution of bipartite entanglement among the $AB$, $AC$, and $BC$ subsystems has been investigated in this configuration \cite{6.3spectatorPhysRevD2024}. Subsequent work has characterized the genuine tripartite entanglement generated in the global $ABC$ state \cite{6.4spectatorPhysRevD2026}.
	The angle-dependent helicity amplitudes raise a further question about measurement outcomes: for a fixed-angle post-selected final state, how strongly are spin-measurement outcomes on $A$ correlated with $B$ and $C$ separately? Entropic uncertainty relations with quantum memory \cite{TwoEn2.2Berta2010,ThreePhysRevLett.103.020402} quantify the uncertainty remaining about these outcomes when a correlated particle is available as quantum side information. For the outgoing electron, $\sigma_z$ distinguishes the two helicities, while $\sigma_x$ and $\sigma_y$ have eigenstates that are coherent superpositions of helicities. The three observables represent alternative measurements on independently prepared scattering events selected within the same angular bin. After Alice announces her measurement basis, Bob or Charlie can use particle $B$ or $C$, respectively, as quantum side information about her outcome. 
	Quantum-memory-assisted entropic uncertainty relations have been investigated in top-quark pair production \cite{5.2EUR} and three-flavor neutrino oscillations \cite{6.4LiJuan}. In the entangled-spectator Bhabha configuration, however, the separate information available to the scattered positron $B$ and spectator electron $C$ about spin-measurement outcomes on $A$ has not been resolved.
		
	We address this question by computing the total uncertainty $U_A$, formed from the conditional uncertainties associated with memories $B$ and $C$, and comparing it with the entanglement entropy $S(\rho_A)$ across the $A|BC$ partition. For the post-selected pure state, $S(\rho_A)$ quantifies entanglement across this partition but does not identify which individual memory is correlated with Alice's outcomes. We therefore examine the separate uncertainties $U_B$ and $U_C$ together with the corresponding outcome-memory mutual-information sums $G_B$ and $G_C$. These information quantities refer to outcomes of specified measurements on $A$, rather than to mutual information between the unmeasured particle states. Comparing them with the outcome uncertainty on $A$ alone distinguishes changes in Alice's outcome distribution from changes in her correlations with each memory.
	The distinction is particularly clear in the high-energy backscattering limit. When the initial $B$-$C$ state is entangled, the spectator $C$ becomes correlated with Alice's spin outcomes, while the scattered positron $B$ carries no outcome information in this limit. The reduced spin state of $C$ is unchanged in the same analytic limit, showing that the outcome information is contained in joint $A$-$C$ correlations rather than in a change of the spectator's local spin statistics. These results concern an idealized fixed-angle post-selected spin state. Their experimental reconstruction would require preparation of the entangled $B$-$C$ input, selection of outgoing events within finite angular bins, and spin-resolved correlation measurements in multiple bases.
 
   To complement the memory-resolved analysis, we employ the entropic complete complementarity relation (CCR) \cite{7.2CCR}. The  CCR has been applied to single- and two-flavor-mode reduced states of a pure three-flavor neutrino state \cite{7.3CCR}. Here, we use it to characterize the helicity-basis coherence, predictability, and von Neumann entropy of $\rho_A$ and $\rho_{AB}$. Because the post-selected $ABC$ state is pure, $S_{\mathrm{vn}}(\rho_A)$ quantifies entanglement across $A|BC$, while $S_{\mathrm{vn}}(\rho_{AB})$ characterizes the mixedness of the interacting pair. The memory-resolved analysis identifies which particle is correlated with Alice's spin-measurement outcomes, and the CCR analysis describes the accompanying changes in the two reduced states.

  The paper is organized as follows. In Sec.~\ref{II}, we introduce the post-selected Bhabha scattering model with an entangled spectator and define the memory-assisted uncertainties, the outcome-memory mutual informations, and the entropic complete complementarity relation. In Sec.~\ref{III}, we first compare the total uncertainty $U_A$ with $S(\rho_A)$. We then examine $U_B$, $U_C$, $G_B$, and $G_C$ to resolve the contributions of the two memories, including their analytic high-energy backscattering limit. Finally, we present the CCR analyses of $\rho_A$ and $\rho_{AB}$. Sec.~\ref{IV} summarizes our conclusions.
	
\section{Theoretical framework}\label{II}

\subsection{Bhabha scattering model}\label{IIA}
	\begin{figure}[t]
	\centering
	\includegraphics[width=0.75\textwidth]{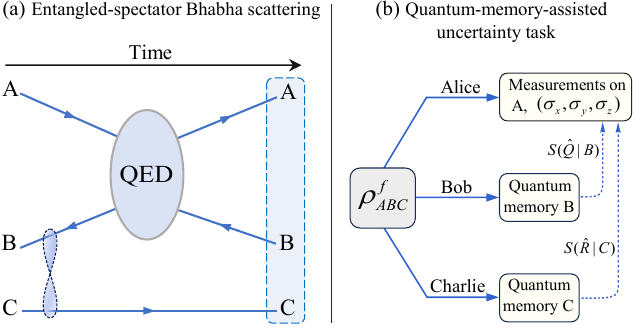}
	\caption{Schematic illustration of entangled-spectator Bhabha scattering and the associated quantum-memory-assisted uncertainty task. Panel (a) shows the initial $B$-$C$ entanglement, the Bhabha scattering between particles $A$ and $B$, and the post-selected three particle state. Panel (b) depicts spin measurements on particle $A$, with particles $B$ and $C$ serving as quantum memories.}
	\label{Fig1}
    \end{figure}	
	First, we briefly review the setup introduced in Refs.~\cite{6.1spectatorPhysRevD.2019,6.2spectatorPhysRevD.2022,6.3spectatorPhysRevD2024}. We consider tree-level QED Bhabha scattering ($e^-e^+ \rightarrow e^-e^+$), where positron $B$ is initially entangled with electron $C$, which acts as a spectator, as shown in Fig.~\ref{Fig1}\textcolor{blue}{(a)}. All calculations are performed in the c.m. frame of particles $A$ and $B$, and we focus on the post-selected three-particle scattering state. We begin by defining the inner product of fermionic states as
	\begin{equation}
		\langle k, a \mid p, b\rangle=2 E_{\mathbf{k}}(2 \pi)^{3} \delta^{(3)}(\mathbf{k}-\mathbf{p}) \delta_{a, b},
		\label{Eq.2.1.1}
	\end{equation}
	where $k$ and $p$ are the 4-momenta, and $a$ and $b$ are the spin indices.
	Suppose that the initial state is
	\begin{equation}
		\begin{split} 
			|i\rangle&=|p_1,a\rangle_A\otimes(\cos\eta|p_2,\uparrow\rangle_B\otimes|q,\uparrow\rangle_C\\
			&+e^{i\beta}\sin\eta|p_2,\downarrow\rangle_B\otimes|q,\downarrow\rangle_C),
			\label{Eq.2.1.2}
		\end{split}
	\end{equation}
	where the parameter $\eta$ determines the initial entanglement between $B$ and $C$. The scattering matrix (S-matrix) is defined as $S=\mathbb{I}+iT$. Here, $\mathbb{I}$ is the identity operator, which characterizes the non-scattering process. By contrast, the operator $T$ is the transition operator describing the genuine Bhabha scattering process mediated by QED interactions, which constitutes the central physical process of our study. After scattering, the final state is
	\begin{equation}
		\begin{split}  	  
			\left|f\right\rangle
			&=|i\rangle+i\sum_{r,s}\int_{\boldsymbol{p}_3,\boldsymbol{p}_4\neq\boldsymbol{p}_1,\boldsymbol{p}_2}\delta^{(4)}(p_1+p_2-p_3-p_4) \\
			&\times[\cos\eta\mathcal{M}(a,\uparrow;r,s)|p_3,r\rangle\otimes|p_4,s\rangle\otimes|q,\uparrow\rangle \\
			&+\mathrm{e}^{i\beta}\sin\eta\mathcal{M}(a,\downarrow;r,s)|p_3,r\rangle\otimes|p_4,s\rangle\otimes|q,\downarrow\rangle],
			\label{Eq.2.1.3}
		\end{split}
	\end{equation}
	where the integral $\int_{\boldsymbol{p}}$ denotes $\int(d^3\boldsymbol{p})/(2E_p(2\pi)^3)$, and $\mathcal{M}(a,\uparrow;r,s)$ represents the scattering amplitude. Since all amplitudes $\mathcal{M}$ have the same dependence on the initial and final momenta, we omit the momentum dependence for simplicity and retain only the spin dependence. According to Eq.~(\ref{Eq.2.1.3}), the density matrix of the final state is
	\begin{equation}
		\rho_{A B C}^{f}=\frac{1}{\mathcal{N}}|f\rangle\langle f|,
		\label{Eq.2.1.4}
	\end{equation}
	where $\mathcal{N}$ is the normalization constant. Taking the partial trace over the subsystems yields
	\begin{equation}
		\text{Tr}_X[\rho{}^f]=\sum_\sigma\int\frac{d^3\mathbf{k}}{(2\pi)^32E_\mathbf{k}}(\mathbb{I}  r\otimes_X\langle k,\sigma|)\rho{}^f(\mathbb{I}  _r\otimes|k,\sigma\rangle_X),
		\label{Eq.2.1.5}
	\end{equation}
	where $\mathbb{I}_r$ is the identity operator on the remaining subspaces, $k$ and $\sigma$ denote the 4-momentum and spin index, respectively, and $X$ denotes the subsystem over which the trace is taken. Employing Eq.~(\ref{Eq.2.1.4}) and the following relations

	\begin{equation}
		2\pi \delta^{(0)}(E_i - E_f) = \int_{-T/2}^{T/2} e^{i(E_i - E_f)t} dt,
		\label{Eq.2.1.6}
	\end{equation}
	\begin{equation}
		(2\pi)^3 \delta^{(3)}(\mathbf{k} - \mathbf{p}) = V \delta_{\mathbf{k},\mathbf{p}},
		\label{Eq.2.1.7}
	\end{equation}
	which imply that  $(2\pi) \delta^{(0)}(0) = T $ and $ (2\pi)^3 \delta^{(3)}(0) = V$, and the normalization constant $\mathcal{N}$ can be readily computed 
	\begin{equation}
		\begin{split}
			\mathcal{N} &= \text{Tr}_A [\text{Tr}_B [\text{Tr}_C [[f \rangle \langle f|]]]] \\
			&= 2E_{p_1} 2E_{p_2} 2E_q V^3 + 2E_q T^2 V^2 \Lambda,
		\end{split}
		\label{Eq.2.1.8}
	\end{equation}
	where
	\begin{equation}
		\begin{split}
			\Lambda= \int \frac{d^3 \mathbf{p_3}}{(2\pi)^3 2 E_{\mathbf{p_3}} 2 E_{\mathbf{p_4}}} \sum_{r,s} (\cos^2 \eta |\mathcal{M}(a, \uparrow; r, s)|^2 
			+ \sin^2 \eta |\mathcal{M}(a, \downarrow; r, s)|^2) |_{\mathbf{p_4} = \mathbf{p_1} + \mathbf{p_2} - \mathbf{p_3}}.
		\end{split}
		\label{Eq.2.1.9}
	\end{equation}
	It should be noted that Eq.~(\ref{Eq.2.1.8}) adopts an additive treatment of the scattering probabilities. A fully unitary $S$-matrix treatment would include forward-scattering interference terms, which cancel the $\Lambda$ term through the optical theorem \cite{scattering4.4}. Since we focus on quantum correlations generated by actual scattering events, we project out the unscattered forward beam and restrict the analysis to the interacting post-selected subensemble, with its normalization specified below.
	
     \subsection{Entanglement entropy, memory-assisted uncertainty, and outcome information}\label{IIB}
	To quantify the correlations between particle $A$ and subsystem $BC$ after scattering, we consider the reduced density matrix
	$\rho_A=\mathrm{Tr}_{BC}(\rho^f_{ABC})$. Since the post-selected final state $\rho^f_{ABC}$ is pure, the entanglement across the $A|BC$ partition is quantified by the von Neumann entropy \cite{7.1Book}
	\begin{equation}
		S(\rho_A)=-\mathrm{Tr}\left(\rho_A\log_2\rho_A\right).
	\end{equation}
	Here, $S(\rho_A)=0$ indicates that $A$ is separable from $BC$, whereas larger values correspond to stronger entanglement across the $A|BC$ partition. 
	
	The entropic uncertainty relation with quantum memory characterizes the uncertainty of incompatible measurements in the presence of correlated memories. For a tripartite system, Renes and Boileau \cite{ThreePhysRevLett.103.020402} proposed
	\begin{equation}
		S(\hat{Q}|B)+S(\hat{R}|C)
		\geq
		\log_2\frac{1}{c(\hat{Q},\hat{R})}
		=q_{MU},
	\end{equation}
	where $B$ and $C$ are the quantum memories. The conditional von Neumann entropy
	$S(\hat{Q}|B)=S(\rho_{\hat{Q}B})-S(\rho_B)$ \cite{7.1Book}
	is evaluated after measuring $\hat{Q}$ on particle $A$, with
	\begin{equation}
		\rho_{\hat{Q}B}
		=
		\sum_i
		\left(
		|\psi_i\rangle\langle\psi_i|\otimes\mathbb{I}_B
		\right)
		\rho_{AB}
		\left(
		|\psi_i\rangle\langle\psi_i|\otimes\mathbb{I}_B
		\right),
	\end{equation}
	where $\{|\psi_i\rangle\}$ are the eigenstates of $\hat{Q}$ and $\mathbb{I}_B$ is the identity operator on $B$. Analogous definitions apply to $S(\hat{R}|C)$ and $\rho_{\hat{R}C}$. The complementarity factor
	\begin{equation}
		c(\hat{Q},\hat{R})
		\equiv
		\max_{i,j}
		|\langle\psi_i|\phi_j\rangle|^2,
	\end{equation}
	is the maximal overlap between the eigenstates of $\hat{Q}$ and $\hat{R}$.
	This relation can be interpreted through a quantum guessing game. As shown in Fig.~\ref{Fig1}\textcolor{blue}{(b)}, Alice, Bob, and Charlie share $\rho_{ABC}$ and hold particles $A$, $B$, and $C$, respectively. Alice measures either $\hat{Q}$ or $\hat{R}$ on $A$ and announces the chosen basis, while Bob and Charlie use their quantum memories to predict her outcome. The relation reflects both measurement complementarity and the restricted shareability of quantum correlations: improved prediction by one memory generally limits the predictive ability of the other for the incompatible measurement.
	
	For the scattering model introduced in Sec.~\ref{IIA}, we consider the Pauli observables $\sigma_x$, $\sigma_y$, and $\sigma_z$, which give
	\begin{equation}
		\begin{split}
			S(\sigma_x|B)+S(\sigma_z|C) &\geq q_{MU},\\
			S(\sigma_y|B)+S(\sigma_x|C) &\geq q_{MU},\\
			S(\sigma_z|B)+S(\sigma_y|C) &\geq q_{MU}.
		\end{split}
	\end{equation}
	Adding these inequalities, we define the total memory-assisted entropic uncertainty as
		\begin{equation}
			\begin{split}
				U_A
				=S(\sigma_x|B)+S(\sigma_z|C)+S(\sigma_y|B)
				+S(\sigma_x|C)+S(\sigma_z|B)+S(\sigma_y|C)
				\geq 3q_{MU}.
			\end{split}
			\label{Eq.2.2.3}
		\end{equation}
		Since any two distinct Pauli bases are mutually unbiased,
		$c(\hat{Q},\hat{R})=1/2$ and hence $q_{MU}=1$, yielding the lower bound $U_A\geq3$. 
	To resolve the contributions associated with the two memories, we regroup the six terms in Eq.~\eqref{Eq.2.2.3} as
		\begin{equation}
			\begin{split}
				U_B &\equiv \sum_{j \in \{x,y,z\}} S(\sigma_j|B),\\
				U_C &\equiv \sum_{j \in \{x,y,z\}} S(\sigma_j|C).
			\end{split}
		\end{equation}
		Here, \(U_B\) and \(U_C\) quantify the residual uncertainty of the measurements on \(A\) when \(B\) and \(C\), respectively, are used as separate quantum memories. Thus, $U_A=U_B+U_C$.  
		To distinguish the uncertainty intrinsic to the measurement outcomes on $A$ from the information encoded in each memory, we introduce the outcome-memory mutual information. Let $\hat{Q}$ denote one of the Pauli observables $\sigma_x$, $\sigma_y$, and $\sigma_z$, and let $Q_A$ denote the classical outcome obtained by measuring $\hat{Q}$ on $A$. With $p_q^{(\hat{Q})}=\mathrm{Tr}(\Pi_q^{(\hat{Q})}\rho_A)$, the intrinsic Shannon entropy of the measurement outcomes is $H(Q_A)=-\sum_q p_q^{(\hat{Q})}\log_2 p_q^{(\hat{Q})}$. For $M \in \{B,C\}$, we define
		\begin{equation}
			\begin{split}
				I(Q_A:M) &= H(Q_A)+S(\rho_M)-S(\rho_{\hat{Q}M})\\
				&= H(Q_A)-S(\hat{Q}|M),
			\end{split}
			\label{eq:outcome-memory-information}
		\end{equation}
		where $\rho_{\hat{Q}M}$ is the post-measurement state defined above for the corresponding memory. The total outcome information encoded in each memory is then
		\begin{equation}
			\begin{split}
				G_B &\equiv \sum_{\hat{Q} \in \{\sigma_x,\sigma_y,\sigma_z\}} I(Q_A:B),\\
				G_C &\equiv \sum_{\hat{Q} \in \{\sigma_x,\sigma_y,\sigma_z\}} I(Q_A:C).
				\label{eq:memory-gains}
			\end{split}
		\end{equation}
		Unlike $U_M$, which measures the outcome uncertainty remaining when
		memory $M$ is available, $G_M$ measures the reduction relative to the
		uncertainty of measuring $A$ without that memory. In the guessing game
		interpretation, $G_M$ quantifies the information about Alice's
		measurement outcomes encoded in the correlations with $M$, when the
		chosen measurement basis is known.
		Defining $H_\Sigma \equiv \sum_{\hat{Q}} H(Q_A)$, we obtain the identities
		\begin{equation}
			\begin{split}		
				U_B &= H_\Sigma-G_B, \qquad U_C=H_\Sigma-G_C,\\
				U_A &= 2H_\Sigma-G_B-G_C, \quad U_B-U_C=G_C-G_B.
			\end{split}
			\label{eq:memory-information-balance}
		\end{equation}
		Although $U_B-U_C=G_C-G_B$ determines the relative outcome information in the two memories at a fixed scattering configuration, the variation of either $U_M$ with the scattering parameters also contains the variation of the intrinsic outcome entropy $H_\Sigma$. 
		For \(M\in\{B,C\}\), \(G_M=H_\Sigma-U_M\) isolates the reduction in outcome uncertainty associated with correlations between Alice’s result and memory \(M\).

		%
	
	\subsection{Entropic complete complementarity relation}
	\label{sec:CCR}
	
	To characterize the reduced states in a specified helicity basis, we use the entropic complete complementarity relation (CCR) \cite{7.2CCR,7.3CCR}. For a density matrix $\rho$ on a $d$-dimensional Hilbert space, let $\rho_{\mathrm{diag}}$ denote the state obtained by removing the off-diagonal elements of $\rho$ in the chosen reference basis. The relative entropy of coherence, entropic predictability, and von Neumann entropy are defined as
	\begin{equation}
		\begin{split}
			C_{\mathrm{re}}(\rho)
			&=S_{\mathrm{vn}}(\rho_{\mathrm{diag}})-S_{\mathrm{vn}}(\rho),\\
			P_{\mathrm{vn}}(\rho)
			&=\log_2 d-S_{\mathrm{vn}}(\rho_{\mathrm{diag}}),\\
			S_{\mathrm{vn}}(\rho)
			&=-\mathrm{Tr}(\rho\log_2\rho).
		\end{split}
	\end{equation}
	These definitions give the exact relation \cite{7.2CCR}
	\begin{equation}
		C_{\mathrm{re}}(\rho)+P_{\mathrm{vn}}(\rho)+S_{\mathrm{vn}}(\rho)
		=\log_2 d.
		\label{eq:CCR_general}
	\end{equation}
	The coherence and predictability depend on the reference basis, whereas the von Neumann entropy does not.
	
	We apply this relation to $\rho_A$ in the single-particle helicity basis $\{|R\rangle,|L\rangle\}$ and to $\rho_{AB}$ in the product basis $\{|RR\rangle,|RL\rangle,|LR\rangle,|LL\rangle\}$. Their dimensions are $d_A=2$ and $d_{AB}=4$, so the corresponding CCR sums are $1$ and $2$, respectively. Because the fixed-angle post-selected state of $ABC$ is pure, $S_{\mathrm{vn}}(\rho_A)$ quantifies entanglement across $A|BC$, while $S_{\mathrm{vn}}(\rho_{AB})=S_{\mathrm{vn}}(\rho_C)$ quantifies entanglement across $AB|C$. In particular, $S_{\mathrm{vn}}(\rho_{AB})$ does not quantify entanglement between $A$ and $B$.

	The CCR tracks the coherence, helicity-basis predictability, and entropy of $\rho_A$ and $\rho_{AB}$ as the scattering parameters vary. Here, predictability means that the populations in the chosen helicity basis depart from a uniform distribution. It does not measure how much information particle $B$ or $C$ contains about the outcomes of spin measurements on $A$; that outcome--memory information is quantified separately by $G_B$ and $G_C$ introduced  in \ref{IIB}.

	\section{results and analysis}\label{III}	
	\subsection{Total memory-assisted uncertainty and \(A|BC\) entanglement}\label{IIIA}	
	\begin{figure*}[t]
		\centering
		\includegraphics[width=1\textwidth]{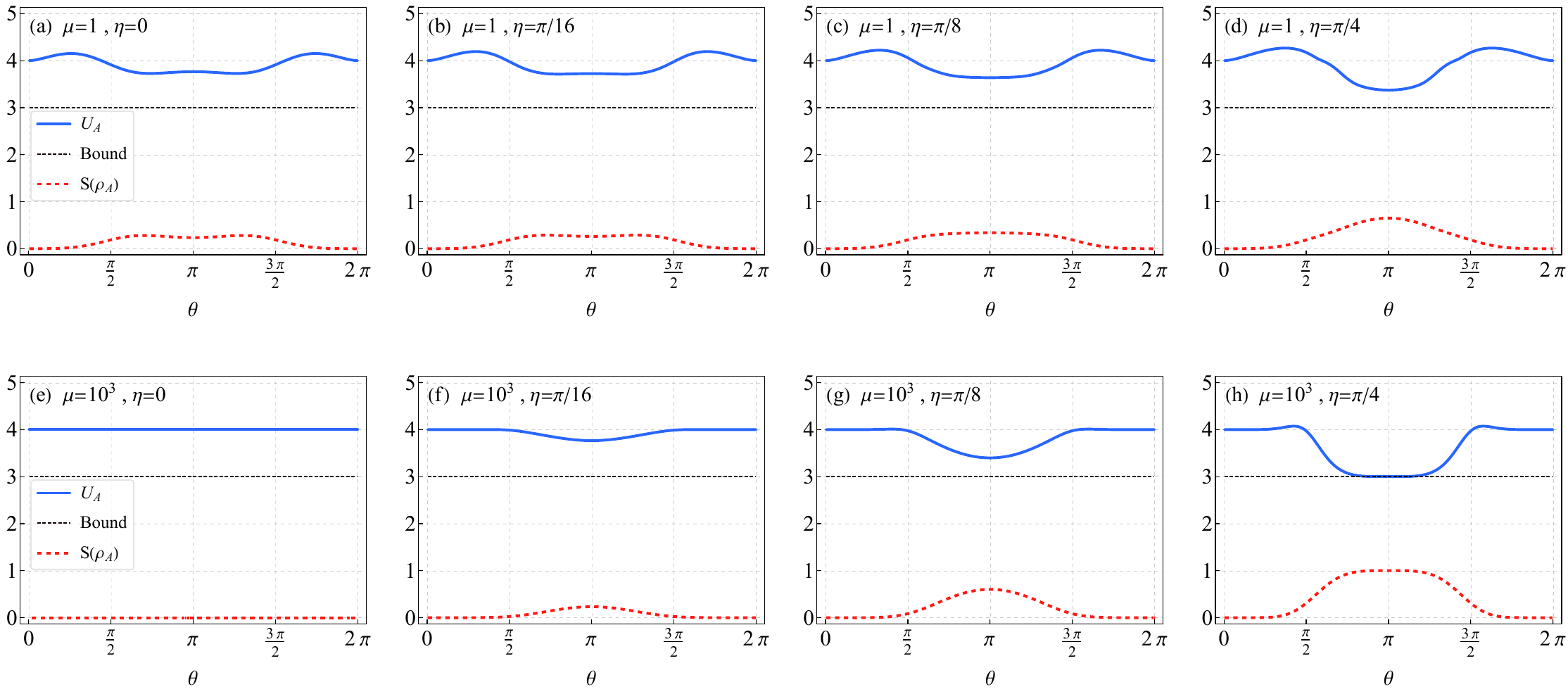}
		\caption{Angular dependence of the memory-assisted entropic uncertainty $U_A$ and the entanglement entropy $S(\rho_A)$ across the $A|BC$ partition. The blue solid, black dashed, and red dashed curves represent $U_A$, its lower bound $U_A=3$, and $S(\rho_A)$, respectively. The upper [(a)--(d)] and lower [(e)--(h)] rows correspond to $\mu=1$ and $\mu=10^3$. From left to right, $\eta=0$, $\pi/16$, $\pi/8$, and $\pi/4$, with $\beta=0$ in all panels.}
		\label{Fig.2}
	\end{figure*}	
	Assuming that the initial state of the system $ABC$ is
	\begin{equation}
		|i\rangle=|R\rangle_{A}\otimes(\cos\eta|R\rangle_{B}|R\rangle_{C}+e^{i\beta}\sin\eta|L\rangle_{B}|L\rangle_{C}),
		\label{Eq.3.1.1}
	\end{equation}
	accordingly, its density matrix can be expressed as
	\begin{equation}
		\begin{aligned}
			\rho_{ABC}^{i} & =\mathrm{cos}^{2}\eta|R\rangle_{A}|R\rangle_{B}|R\rangle_{CC}\langle R|_{B}\langle R|_{A}\langle R|\\
			&+e^{-i\beta}\sin\eta\cos\eta|R\rangle_{A}|R\rangle_{B}|R\rangle_{CC}\langle L|_{B}\langle L|_{A}\langle R| \\
			& +e^{i\beta}\sin\eta\cos\eta|R\rangle_{A}|L\rangle_{B}|L\rangle_{CC}\langle R|_{B}\langle R|_{A}\langle R|\\
			&+\mathrm{sin}^{2}\eta|R\rangle_{A}|L\rangle_{B}|L\rangle_{CC}\langle L|_{B}\langle L|_{A}\langle R|.
		\end{aligned}
		\label{Eq.3.1.2}
	\end{equation}
	In the c.m. frame of the interacting pair $AB$, we choose the incoming electron $A$ and positron $B$ to move along the $+z$ and $-z$ directions, respectively. With $\omega=\sqrt{m_e^2+|\vec p|^2}$, their 4-momenta are
	\begin{equation}
		\begin{split}
			&p_1 = (\omega, 0, 0, |\vec{p}|),\\
			&p_2 = (\omega, 0, 0, -|\vec{p}|).
			\label{Eq.55}
		\end{split}
	\end{equation}
	For the final state, we parametrize the outgoing electron and positron momenta in the $xz$ plane as		
	\begin{equation}
		\begin{split}
			&p_3 = (\omega, |\vec{p}|\sin\theta, 0, |\vec{p}|\cos\theta),\\
			&p_4 = (\omega, -|\vec{p}|\sin\theta, 0, -|\vec{p}|\cos\theta),
			\label{Eq.66}
		\end{split}
	\end{equation}
	here, $\theta\in(0,2\pi)$ is an oriented angle in the $xz$ plane.

	To characterize the scattering energy conveniently, we introduce the dimensionless scattering momentum $\mu=|\mathbf{p}|/m_e$, where $|\mathbf{p}|$ denotes the magnitude of the incoming momentum in the c.m. reference frame and $m_e$ is the electron mass. From Eqs. (\ref{Eq.2.1.3}) and (\ref{Eq.2.1.4}), one can obtain the density matrix of the full final state. In the present work, however, we are interested in the quantum correlations generated by actual scattering events. Therefore, we restrict our analysis to particles scattered at nonzero angles. 
After selecting the outgoing momenta corresponding to a fixed scattering angle and projecting out the unscattered contribution, the unnormalized spin state is
\begin{equation}
	|\widetilde\Psi_f\rangle
	=\cos\eta\sum_{r,s}\mathcal M(RR;rs)|rsR\rangle
	+e^{i\beta}\sin\eta\sum_{r,s}\mathcal M(RL;rs)|rsL\rangle.
	\label{Eq.3.1.26}
\end{equation}
Since the spectator states in the two terms are orthogonal, its squared norm is
\begin{equation}
\mathcal Z(\theta,\mu;\eta)
=\cos^2\eta\sum_{r,s}|\mathcal M(RR;rs)|^2
+\sin^2\eta\sum_{r,s}|\mathcal M(RL;rs)|^2.
\label{Eq.3.1.27}
\end{equation}
The normalized fixed-angle spin density matrix is therefore
\begin{equation}
\rho^f_{ABC}(\theta,\mu)
=\frac{|\widetilde\Psi_f\rangle
	\langle\widetilde\Psi_f|}
{\mathcal Z(\theta,\mu;\eta)}.
\label{Eq.3.1.3}
\end{equation}
	The scattering amplitudes  $\mathcal{M}$ depend on the scattering angle \(\theta\) and the dimensionless momentum \(\mu\), their explicit expressions are given in Appendix \ref{A}.
	By using Eq. (\ref{Eq.3.1.3}), the reduced density matrices $\rho_A^{f}$, $\rho_B^{f}$, $\rho_C^{f}$, as well as $\rho_{AB}^{f}$, $\rho_{AC}^{f}$, and $\rho_{BC}^{f}$, can be obtained. The explicit forms of these reduced density matrices are presented in Appendix \ref{B}. 
	
\begin{figure}[t]
    	\centering
    	\includegraphics[width=0.75\textwidth]{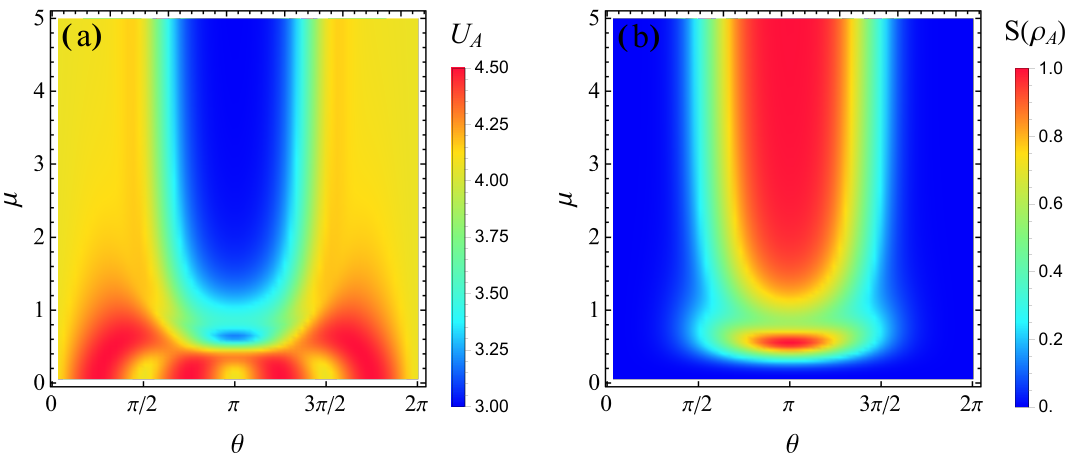}
    	\caption{Density plots of (a) the total memory-assisted uncertainty $U_A$ and (b) the entanglement entropy $S(\rho_A)$ in the $(\theta,\mu)$ plane for $\eta=\pi/4$ and $\beta=0$. The displayed momentum range is $0<\mu\leq5$. The color bars are scaled independently: dark blue in panel (a) indicates values of $U_A$ close to its lower bound of $3$, whereas red in panel (b) indicates values of $S(\rho_A)$ close to its maximum of $1$.}
    	\label{Fig.3}
    \end{figure}

    To examine how the scattering configuration affects the total memory-assisted uncertainty, we compare $U_A=U_B+U_C$ with the entanglement entropy $S(\rho_A)$ across the $A|BC$ partition. Figure~\ref{Fig.2} shows their angular dependence for different initial $B$-$C$ entanglement weights $\eta$ and incident momenta $\mu$. The blue solid and red dashed curves represent $U_A$ and $S(\rho_A)$, respectively, while the horizontal black dashed line marks the lower bound $U_A=3$. The calculated uncertainty never falls below this bound. In the backscattering region, an increase in $S(\rho_A)$ is generally accompanied by a decrease in $U_A$, although the two quantities do not vary oppositely at every scattering angle.
    For the lower-momentum regime $\mu=1$, shown in Figs.~\ref{Fig.2}(a)-(d), both quantities vary smoothly with $\theta$. At $\eta=0$, the spectator $C$ is initially uncorrelated with the scattering pair; the nonzero $S(\rho_A)$ at intermediate angles therefore reflects entanglement generated between $A$ and $B$. As $\eta$ increases, $S(\rho_A)$ becomes larger near $\theta\simeq\pi$, while $U_A$ decreases in the same region. This trend is most evident for $\eta=\pi/4$, although the uncertainty reduction remains moderate compared with that in the high-momentum regime.
    For $\mu=10^3$, Figs.~\ref{Fig.2}(e)-(h) show a more pronounced angular dependence. When $\eta=0$, $S(\rho_A)$ remains close to zero and $U_A$ stays near $4$ throughout the displayed angular range. For nonzero $\eta$, a peak in $S(\rho_A)$ and a corresponding dip in $U_A$ develop around $\theta=\pi$. Their magnitudes increase with $\eta$. At $\eta=\pi/4$, the entropy approaches its maximal value of $1$, while the total uncertainty nearly saturates its lower bound of $3$ near backscattering.
    Taking the high-energy backscattering limit of the helicity amplitudes given in the Appendix \ref{A}, we obtain the following normalized post-selected final state
    \begin{equation}
    	|\Psi_f\rangle
    	\xrightarrow[\mu\to\infty]{\theta=\pi}
    	|R\rangle_B
    	\left(
    	\cos\eta\,|RR\rangle_{AC}
    	+e^{i\beta}\sin\eta\,|LL\rangle_{AC}
    	\right),
    	\label{Eq.3.1.6}
    \end{equation}
    \begin{equation}
    	S(\rho_A)\longrightarrow h_2(\sin^2\eta),
    	\qquad
    	U_A\longrightarrow 4-S(\rho_A),
    \end{equation}
    where \(h_2(p)=-p\log_2p-(1-p)\log_2(1-p)\) represents binary entropy. Thus, $U_A+S(\rho_A)\to4$ in this particular limit; the same relation need not hold at finite momentum or away from backscattering.     
    It is worth noting that changing \(\beta\) is equivalent to a local phase rotation of particle \(C\). This leaves the entropy-based quantities considered here unchanged, so we set \(\beta=0\) in all plots.
    To examine how the angular pattern evolves over a finite range of incident momenta, we next consider the density plots in Fig.~\ref{Fig.3}.

    Figure~\ref{Fig.3} extends the fixed-momentum angular scans of Fig.~\ref{Fig.2} to the $(\theta,\mu)$ plane at $\eta=\pi/4$. Within the displayed momentum range, a region of reduced $U_A$ is centered around backscattering. Its angular extent grows as $\mu$ increases through the lower part of the plotted interval. Fig.~\ref{Fig.3}(b) shows enhanced $S(\rho_A)$ over much of the same region.
	The density plots also reveal a narrower feature near $\mu\simeq0.6$ and $\theta\simeq\pi$, where $U_A$ is reduced and $S(\rho_A)$ is enhanced. This feature is not resolved by the two momentum values used in Fig.~\ref{Fig.2} and illustrates that the dependence on $\mu$ need not be monotonic at fixed angle. The present plots cover $\mu\leq5$; the behavior at $\mu=10^3$ is shown separately in Fig.~\ref{Fig.2} and is addressed by the high-energy backscattering result above.
    Figs.~\ref{Fig.2} and \ref{Fig.3} establish where reduced total uncertainty and enhanced $A|BC$ entanglement occur together in the parameter ranges studied. They do not, by themselves, identify which individual memory is correlated with the measurement outcomes on $A$: $S(\rho_A)$ concerns the combined $A|BC$ partition, while changes in $U_A$ also contain changes in Alice's outcome entropy. We therefore turn to $U_B$ and $U_C$, together with $G_B$ and $G_C$, to resolve the contributions associated with the two memories.

	\subsection{Memory-resolved entropic uncertainty and outcome information}	
	
\begin{figure*}[t]
	\centering
	\includegraphics[width=0.75\textwidth]{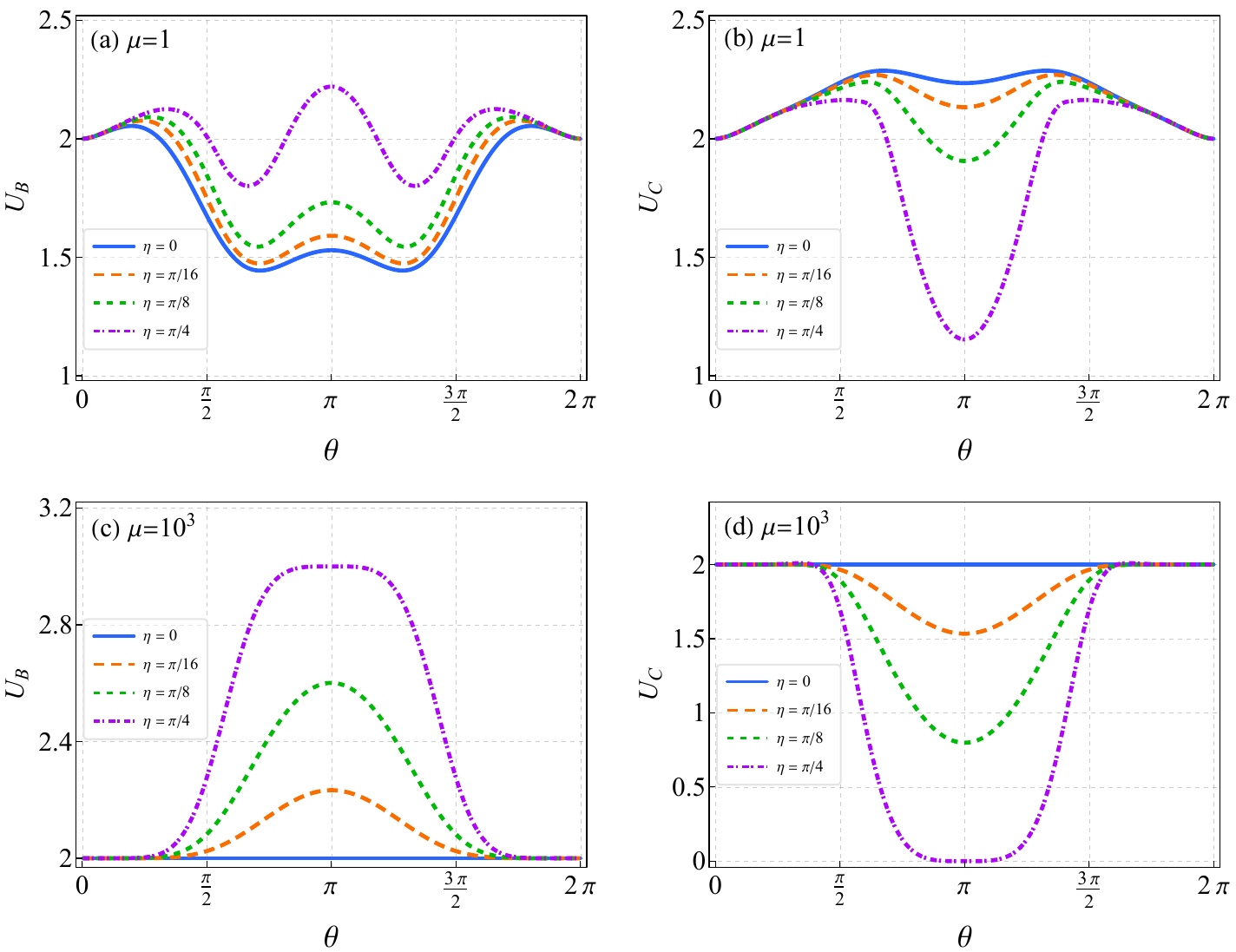}
	\caption{
		\textcolor{black}{Angular dependence of the memory-resolved entropic uncertainties $U_B$ and $U_C$ in Bhabha scattering with an entangled spectator. Panels (a) and (b) show $U_B$ and $U_C$, respectively, for $\mu=1$; panels (c) and (d) show the corresponding results for $\mu=10^3$. The blue solid, orange dashed, green dashed, and purple dash-dotted curves correspond to $\eta=0,\pi/16,\pi/8,\pi/4$, respectively. The initial phase is fixed at $\beta=0$.}}
	\label{fig:UBUC}
\end{figure*}

	Figure~\ref{fig:UBUC} resolves the total uncertainty $U_A=U_B+U_C$ into the contributions associated with the two separate memories. In the lower-momentum regime, $\mu=1$, both contributions vary smoothly with the scattering configuration. For $\eta=0$, $U_B$ develops two angular minima, whereas $U_C$ shows an approximately opposite trend. Since the spectator $C$ is then uncorrelated with $A$, $G_C=\sum_Q I(Q_A:C)=0$ and $U_C=\sum_Q H(Q_A)$. Its angular variation in this case therefore reflects the changing uncertainty of the outcomes on $A$, rather than information acquired by $C$. As $\eta$ increases, $U_B$ rises and $U_C$ falls near $\theta=\pi$. For $\eta=\pi/4$, $U_C<U_B$ in this region; equivalently, $G_C>G_B$, because $U_B-U_C=G_C-G_B$. The decrease in $U_C$ outweighs the increase in $U_B$, producing the reduction in $U_A$ observed in Fig.~\ref{Fig.2}.
	The separation is more pronounced at $\mu=10^3$. When $\eta=0$, both $U_B$ and $U_C$ remain close to 2. For nonzero initial $B$-$C$ entanglement, a peak develops in $U_B$ around the backscattering configuration $\theta=\pi$, accompanied by a dip in $U_C$. At $\eta=\pi/4$, these quantities approach 3 and 0, respectively. According to Eqs. (\ref{eq:memory-gains})-(\ref{eq:memory-information-balance}) and Eq. (\ref{Eq.3.1.6}), this behavior agrees with the high-energy backscattering limit
	\begin{equation}
		\begin{split}
			U_B&\longrightarrow 2+h_2(\sin^2\eta),\\
			U_C&\longrightarrow 2-2h_2(\sin^2\eta),
		\end{split}
	\end{equation}
	\begin{equation}
		\begin{split}
			G_B&\to 0,\\
			G_C&\to3h_2(\sin^2\eta),
		\end{split}
		\label{Eq.29}
	\end{equation}
	the scattered positron $B$ carries no information about the chosen spin outcome on $A$, whereas the spectator $C$ provides the dominant memory contribution. Thus, the reduction of $U_A$ near backscattering is driven primarily by the correlations between $A$ and $C$.

\begin{figure*}[t]
	\centering
	\includegraphics[width=0.75\textwidth]{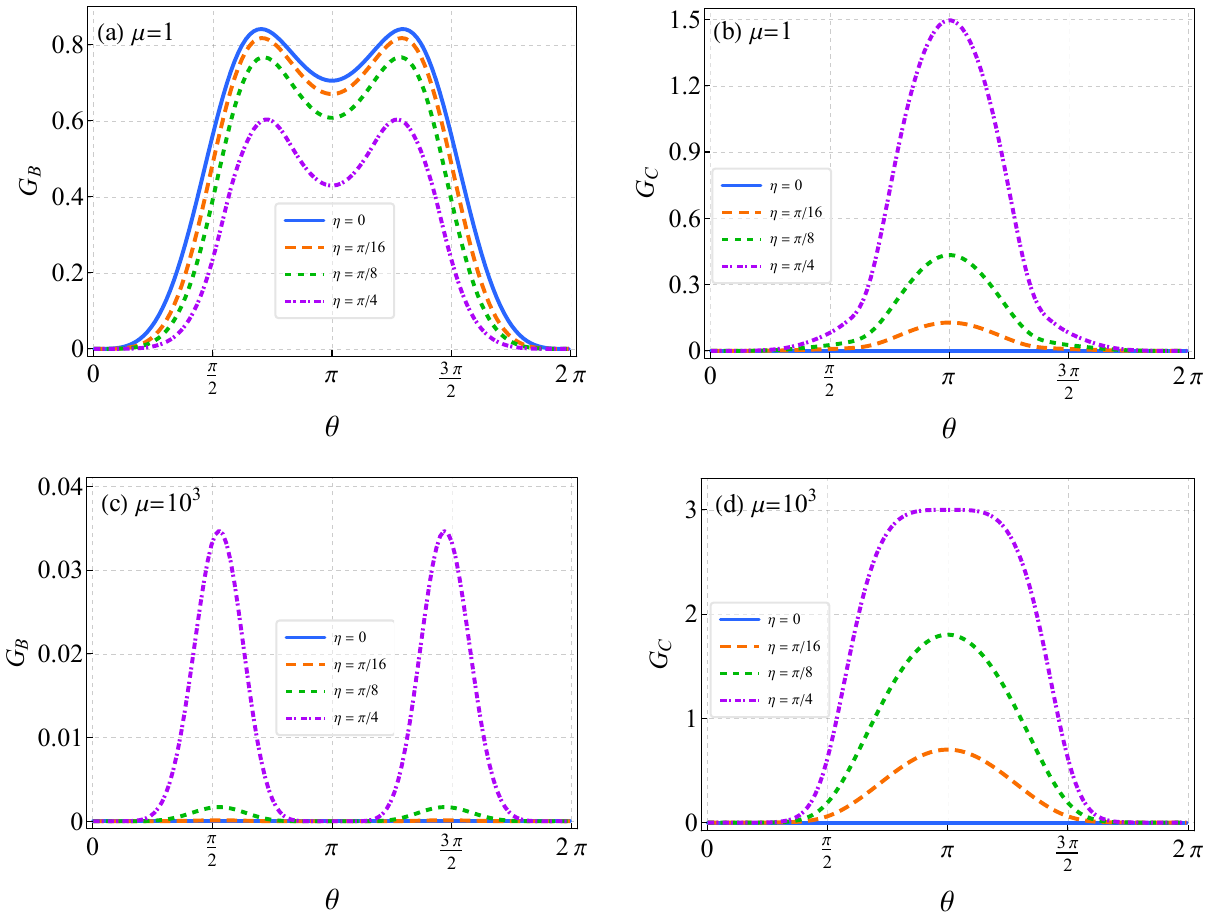}
	\caption{Measurement-outcome information available in the separate  quantum memories after fixed-angle Bhabha scattering. The quantities $G_M=\sum_{Q\in\{\sigma_x,\sigma_y,\sigma_z\}}I(Q_A:M)$, with $M=B,C$, measure the reduction in uncertainty about the outcome of a measurement on $A$ when memory $M$ is available. Panels (a) and (b) show $G_B$ and $G_C$, respectively, for $\mu=1$; panels (c) and (d) show the corresponding results for $\mu=10^3$. The other parameters are the same as in Fig.~\ref{fig:UBUC}.}
	\label{fig:GBGC}
\end{figure*}

  Figure~\ref{fig:GBGC} separates the information about the measurement  outcomes of $A$ that is present in the two memories. For either  $M=B$ or $C$, the relation $U_M=\sum_{Q}H(Q_A)-G_M$ shows that $G_M$ isolates the memory-dependent reduction in uncertainty from the entropy of the outcomes measured on  $A$. 
  In the lower-momentum regime, Fig.~\ref{fig:GBGC}(a) shows that $G_B$ is appreciable when $\eta=0$ and generally decreases as $\eta$  increases. By contrast, $G_C$ vanishes at $\eta=0$ and grows with  $\eta$, reaching its largest values near backward scattering   [Fig.~\ref{fig:GBGC}(b)]. Thus, the two memories do not acquire the  same amount of information about $A$: increasing the initial $B$-$C$ entanglement enhances the information present in $C$ while  reducing that present in $B$ over the angular region where the curves  are most pronounced.  
  This separation becomes particularly clear for $\mu=10^3$. Figure~\ref{fig:GBGC}(c) shows that $G_B$ exhibits small, localized peaks near the scattering angles $\theta = \pi/2$ and $3\pi/2$, whereas $G_C$ becomes dominant in the  backward-scattering region [Fig.~\ref{fig:GBGC}(d)]. The distinction  can be understood from the limiting postselected state at  $\theta=\pi$.  According to Eq. (\ref{Eq.29}),  $G_C$ approaches three bits for $\eta=\pi/4$, as seen in Fig.~\ref{fig:GBGC}(d). At the same time
  \begin{equation}
  	\begin{split}
  		\sum_Q H(Q_A)\to 2+h_2(\sin^2\eta), \quad
  		U_A\to 4-h_2(\sin^2\eta).
  	\end{split}
  \end{equation} 
  Therefore, the decrease of the total uncertainty near backward scattering is associated primarily with information available in the spectator memory $C$, even though the outcomes of measurements on $A$ become intrinsically more uncertain.

 \subsection{CCR analysis of subsystem $A$}
 
 	Having resolved the outcome information associated with memories $B$ and $C$, we now examine the accompanying changes in the reduced state of the measured particle $A$. In the helicity basis $\{|R\rangle,|L\rangle\}$, the entropic complete complementarity relation for this two-dimensional subsystem is
 	\begin{equation}
 		C_{\mathrm{re}}(\rho_A)+P_{\mathrm{vn}}(\rho_A)+S_{\mathrm{vn}}(\rho_A)=1.
 		\label{eq:CCR_A}
 	\end{equation}
 	\begin{figure*}[t]
 		\centering
 		\includegraphics[width=1\textwidth]{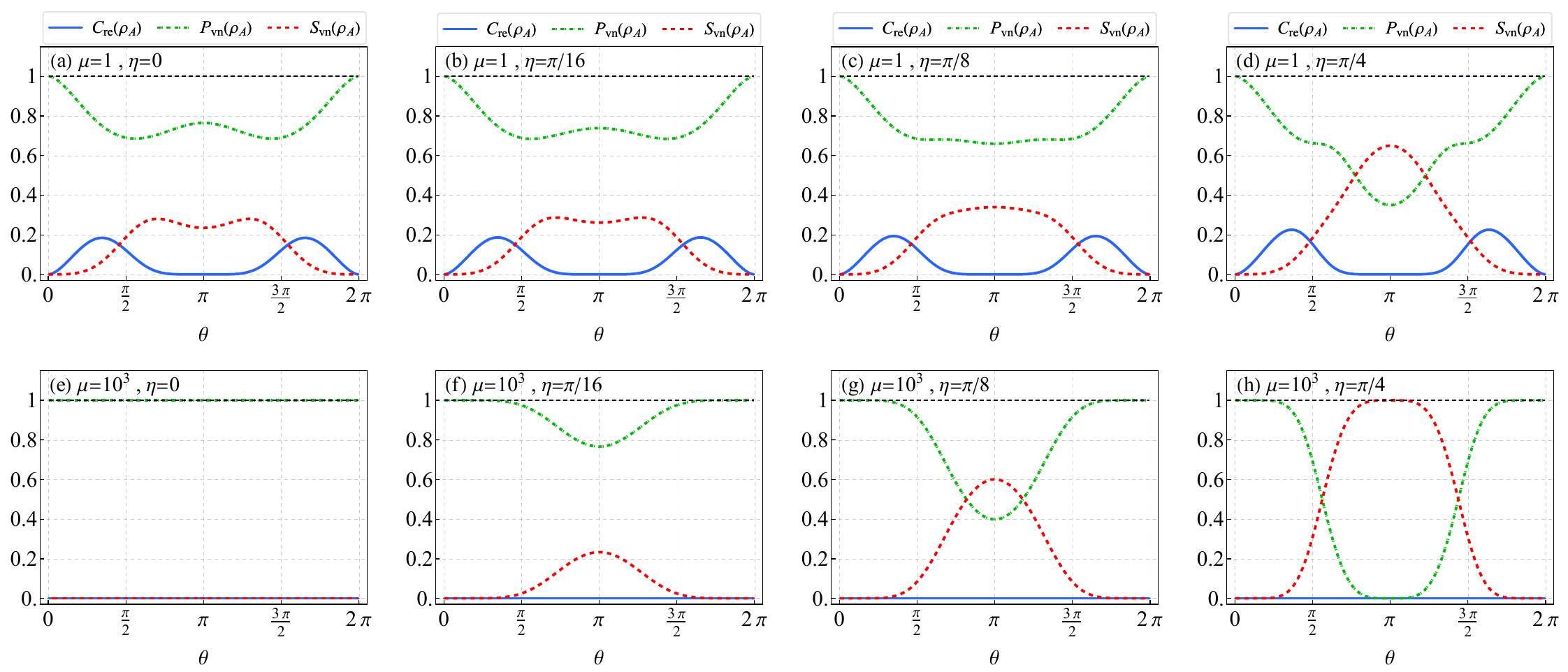}
 		\caption{Angular dependence of the entropic complete complementarity relation components of the reduced state $\rho_A$ in the helicity basis. The blue solid, green dash-dotted, and red dashed curves show $C_{\mathrm{re}}(\rho_A)$, $P_{\mathrm{vn}}(\rho_A)$, and $S_{\mathrm{vn}}(\rho_A)$, respectively; the black dashed line shows their sum. Panels (a)-(d) correspond to $\mu=1$, and panels (e)-(h) to $\mu=10^3$. In each row, $\eta=0$, $\pi/16$, $\pi/8$, and $\pi/4$ from left to right. The phase is $\beta=0$ throughout.}
 		\label{Fig.4}
 	\end{figure*}	
 	Because the fixed-angle post-selected state of $ABC$ is pure, $S_{\mathrm{vn}}(\rho_A)=S(\rho_A)$ quantifies entanglement across the $A|BC$ partition. Equation~\eqref{eq:CCR_A} specifies how this entropy is balanced by helicity-basis coherence and predictability in the local state of $A$; it does not distinguish the information available to the two memories.
 
 Figure~\ref{Fig.4} displays the three components for the parameters used in Fig.~\ref{Fig.2}. The horizontal black line remains at unity, providing a check of their numerical evaluation. For $\mu=1$, Figs.~\ref{Fig.4}(a)-(d) show appreciable coherence at intermediate side angles, while the coherence is small near $\theta\simeq\pi$. At $\eta=0$, the spectator $C$ is uncorrelated with the scattering pair, so the nonzero $S_{\mathrm{vn}}(\rho_A)$ reflects entanglement generated between $A$ and $B$. As $\eta$ increases, $S_{\mathrm{vn}}(\rho_A)$ grows near backscattering and $P_{\mathrm{vn}}(\rho_A)$ decreases there. In this angular region, the change in the CCR balance is therefore dominated by the variations of entropy and local helicity predictability.
 
The pattern is more pronounced at $\mu=10^3$, as shown in Figs.~\ref{Fig.4}(e)-(h). For $\eta=0$, $P_{\mathrm{vn}}(\rho_A)$ stays close to unity, whereas both $C_{\mathrm{re}}(\rho_A)$ and $S_{\mathrm{vn}}(\rho_A)$ are nearly zero. When $\eta>0$, a peak in $S_{\mathrm{vn}}(\rho_A)$ develops around $\theta=\pi$ together with a dip in $P_{\mathrm{vn}}(\rho_A)$. At $\eta=\pi/4$, the entropy approaches unity and the predictability approaches zero over a broad central angular region. The coherence remains close to zero across the displayed angles, so the CCR near backscattering is well approximated by $P_{\mathrm{vn}}(\rho_A)+S_{\mathrm{vn}}(\rho_A)\simeq1$.
 
 This behavior can be checked analytically at $\theta=\pi$ in the limit $\mu\to\infty$. The reduced state approaches
 	\begin{equation}
 		\rho_A\longrightarrow
 		\cos^2\eta\,|R\rangle\langle R|
 		+\sin^2\eta\,|L\rangle\langle L|,
 	\end{equation}
 	which is diagonal in the reference basis. Writing $h=h_2(\sin^2\eta)$, its CCR components therefore satisfy
 	\begin{equation}
 		\begin{split}
 		 C_{\mathrm{re}}(\rho_A)&\longrightarrow0,\\
 		P_{\mathrm{vn}}(\rho_A)&\longrightarrow1-h,\\
 		S_{\mathrm{vn}}(\rho_A)&\longrightarrow h.
 		\end{split}
 	\end{equation}
 	These expressions account for the limiting values shown in Figs.~\ref{Fig.4}(e)-(h).
    The CCR thus identifies the local-state change accompanying the decrease of $U_A$: near high-energy backscattering, greater $A|BC$ entanglement coincides mainly with lower helicity predictability, rather than a substantial change in local coherence. The memory-resolved quantities give the distinct explanation of the uncertainty reduction. 
 
 \subsection{CCR analysis of the interacting subsystem $AB$}
 
  \begin{figure*}[t]
 	\centering
 	\includegraphics[width=0.98\textwidth]{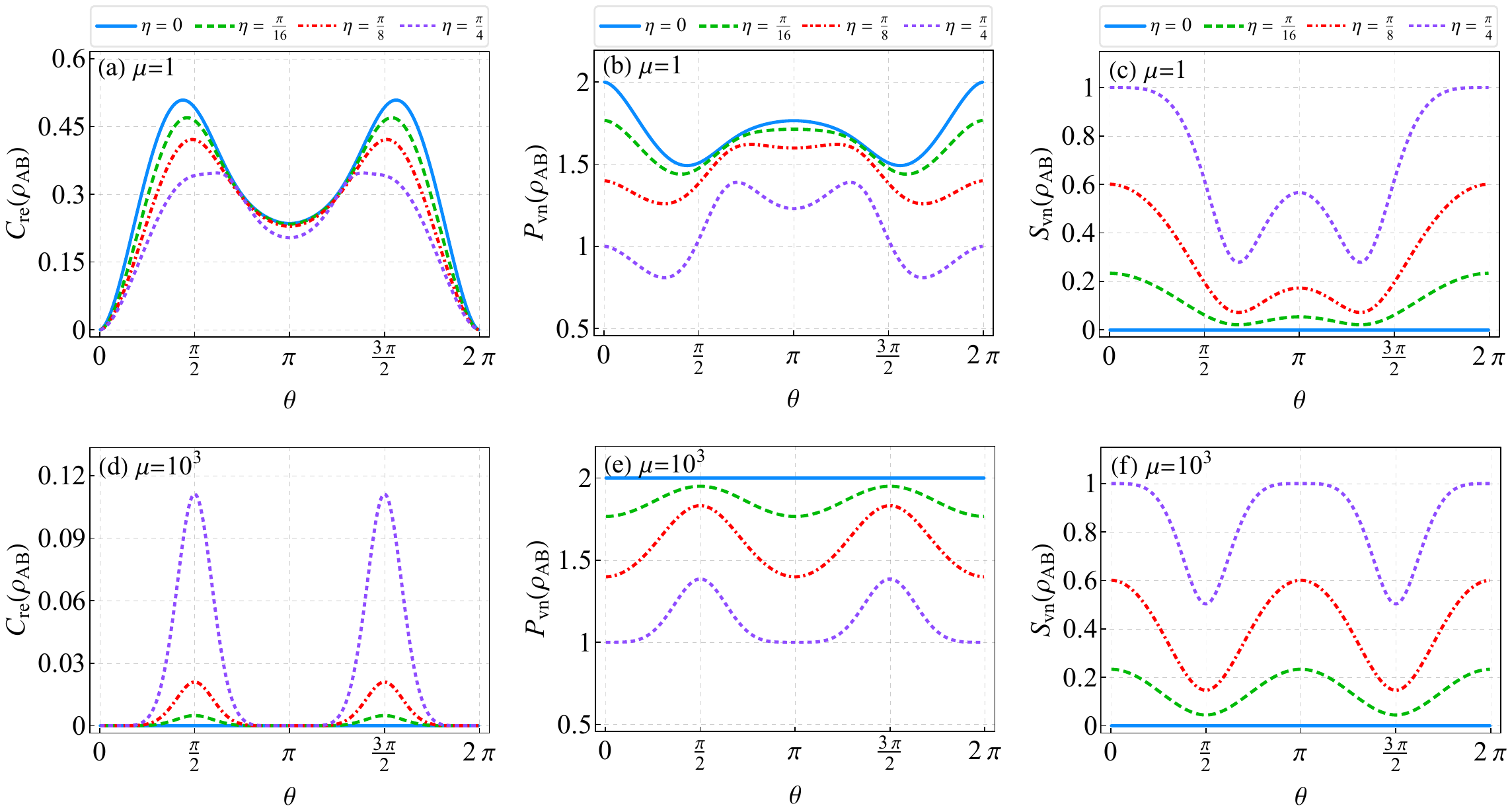}
 	\caption{Angular dependence of the entropic CCR components of the interacting subsystem $AB$. Panels (a)-(c) correspond to $\mu=1$, and panels (d)-(f) to $\mu=10^3$. From left to right, the panels show $C_{\mathrm{re}}(\rho_{AB})$, $P_{\mathrm{vn}}(\rho_{AB})$, and $S_{\mathrm{vn}}(\rho_{AB})$. The blue solid, green dashed, red dash-dotted, and purple dotted curves correspond to $\eta=0$, $\pi/16$, $\pi/8$, and $\pi/4$, respectively. The phase is $\beta=0$ throughout. The constant sum of the three components is omitted.}
 	\label{Fig.5}
 \end{figure*}

 We next examine the reduced state $\rho_{AB}$ of the interacting
 electron-positron pair. Its product helicity basis is
 $\{|RR\rangle,|RL\rangle,|LR\rangle,|LL\rangle\}$,
 and its Hilbert-space dimension is $d_{AB}=4$.
 Equation~(\ref{eq:CCR_general}) then gives
 \begin{equation}
 	C_{\mathrm{re}}(\rho_{AB})
 	+P_{\mathrm{vn}}(\rho_{AB})
 	+S_{\mathrm{vn}}(\rho_{AB})=2.
 	\label{eq:CCR_AB}
 \end{equation}
 Here, $P_{\mathrm{vn}}(\rho_{AB})$ characterizes predictability in the joint helicity basis; it should not be interpreted as Bob's ability to predict Alice's measurement outcome. Since the fixed-angle post-selected state of $ABC$ is pure, $S_{\mathrm{vn}}(\rho_{AB})=S_{\mathrm{vn}}(\rho_C)$ quantifies entanglement across the $AB|C$ partition. 
 
 Figure~\ref{Fig.5} shows the three components for different initial $B$-$C$ entanglement weights. At $\eta=0$, particle $C$ remains separable from $AB$, so $\rho_{AB}$ is pure and $S_{\mathrm{vn}}(\rho_{AB})=0$ at the plotted angles. For $\mu=1$, Figs.~\ref{Fig.5}(a)-(c) show that increasing $\eta$ generally suppresses the side-angle peaks of $C_{\mathrm{re}}(\rho_{AB})$ and lowers $P_{\mathrm{vn}}(\rho_{AB})$, while $S_{\mathrm{vn}}(\rho_{AB})$ increases. In this regime, the change in the CCR balance is associated mainly with lower joint helicity-basis predictability, together with a smaller reduction in coherence.

 At $\mu=10^3$, the coherence responds differently to $\eta$. It is nearly zero for $\eta=0$, but increasing the initial $B$-$C$ entanglement produces small peaks near the side angles, as shown in Fig.~\ref{Fig.5}(d). These peaks remain away from the backscattering region, where the decrease in $U_A$ is most pronounced. Figs.~\ref{Fig.5}(e) and \ref{Fig.5}(f) show that larger $\eta$ generally lowers $P_{\mathrm{vn}}(\rho_{AB})$ and raises $S_{\mathrm{vn}}(\rho_{AB})$. Thus, the coherence of the interacting pair has a different momentum dependence from its predictability and entropy.
 
 The high-energy backscattering limit clarifies the interpretation of the mixedness. In this limit, $\rho_{AB}$ factorizes into $\rho_A$ and a pure helicity state of $B$, while $S_{\mathrm{vn}}(\rho_{AB})\to h_2(\sin^2\eta)$. In particular, at $\eta=\pi/4$ the entropy of $\rho_{AB}$ approaches $1$ even though $A$ and $B$ are unentangled in this limiting state. The entropy reflects correlations across $AB|C$, consistently with the memory-resolved finding that the outcome information resides in $C$ rather than $B$.
 
 The CCR analyses of $\rho_A$ and $\rho_{AB}$ therefore describe different reduced-state properties accompanying the scattering. The former resolves the local helicity-basis balance of the measured particle; the latter shows how initial spectator entanglement changes the coherence, joint-basis predictability, and mixedness of the interacting pair. The separate outcome-memory quantities $G_B$ and $G_C$ determine which particle carries information about measurements on $A$.

	\section{Conclusion} \label{IV}
	
	In this work, we studied memory-assisted entropic uncertainty and complete complementarity in fixed-angle post-selected Bhabha scattering with an entangled spectator. Taking the outgoing electron $A$ as the measured subsystem, we compared the total uncertainty $U_A$ with the entanglement entropy $S(\rho_A)$ across the $A|BC$ partition. We then resolved $U_A$ into the separate uncertainties $U_B$ and $U_C$ and evaluated the outcome--memory information quantities $G_B$ and $G_C$. This separation distinguishes changes in the outcome uncertainty of $A$ from changes in its correlations with each memory.
	
    In the parameter regions examined, $U_A$ and $S(\rho_A)$ exhibit an anticorrelation that is particularly pronounced near high-energy backscattering. Resolving the two memories shows that a decrease in $U_A$ does not imply an improvement in both memories. At $\mu\to\infty$ and $\theta=\pi$, we find $G_B\to0$ and $G_C\to3h_2(\sin^2\eta)$. Thus, the scattered positron $B$ carries no information about Alice's measurement outcomes according to this measure, whereas the spectator $C$ carries increasing outcome information as the initial $B$--$C$ entanglement grows. In the same limit, the increase in $U_B$ reflects a more random outcome distribution on $A$, even though $G_B$ remains zero. The net reduction in $U_A$ is driven by the growing outcome correlations with $C$.
	
	The CCR analysis provides a complementary description of the reduced states. Near high-energy backscattering, the increase in $S_{\mathrm{vn}}(\rho_A)$ is accompanied mainly by a decrease in the local helicity predictability $P_{\mathrm{vn}}(\rho_A)$, while $C_{\mathrm{re}}(\rho_A)$ remains comparatively small. For the interacting pair, the CCR tracks changes in the coherence, predictability, and mixedness of $\rho_{AB}$; because the post-selected global state is pure, $S_{\mathrm{vn}}(\rho_{AB})=S_{\mathrm{vn}}(\rho_C)$.
	
   These results concern an idealized fixed-angle post-selected state; experimental reconstruction would require preparation of an entangled $B$--$C$ input and spin-resolved measurements in multiple bases. Within this setting, they show how Bhabha scattering distributes information about the outgoing electron's spin outcomes between the scattered positron and the spectator, extending the characterization of relativistic QED scattering beyond entanglement measures alone.

	\begin{acknowledgements}	
		This work was supported by the National Natural Science Foundation of China (Grants No. 12475009, and No. 12075001),  Anhui Province Science and Technology Innovation Project (Grant No. 202423r06050004), Anhui Provincial Natural Science Foundation (Grant No.~ 2508085ZD001), Anhui Provincial University Scientific Research Major Project (Grant No. 2024AH040008), and Anhui Provincial Department of Industry and Information Technology (Grant No. JB24044).  
	\end{acknowledgements}
	
	\appendix
	\begin{widetext}
		\section{Bhabha scattering amplitudes}\label{A}
		{The Bhabha scattering amplitude calculated in the c.m. reference frame of particles $A$ and $B$ is written as
			\begin{equation}
				\mathcal{M}_{\mathrm{Bhabha}}=e^2\left(\bar{v}(b,p_2)\gamma^\mu u(a,p_1)\frac{1}{(p_1+p_2)^2}\bar{u}(r,p_3)\gamma_\mu v(s,p_4)-\bar{v}(b,p_2)\gamma^\mu v(s,p_4)\frac{1}{(p_3-p_1)^2}\bar{u}(r,p_3)\gamma_\mu u(a,p_1)\right),
			\end{equation}
			where $a, b, r, s$ are the spin indices. Using the kinematic configuration in Eqs. (\ref{Eq.55})-(\ref{Eq.66}), the explicit expressions  for the polarized amplitudes are given by 	
			
			\begin{equation}\mathcal{M}(RR;RR)=\mathcal{M}(LL;LL)=\frac{(2+11\mu^2+8\mu^4+2\cos\theta+\mu^2\cos2\theta)\csc^2(\frac{\theta}{2})}{4\mu^2(1+\mu^2)},\end{equation}
			\begin{equation}\mathcal{M}(RR;_{LR}^{RL})=-\mathcal{M}(LL;_{LR}^{RL})=-\frac{(1+\mu^2\cos\theta)\cot(\frac{\theta}{2})}{\mu^2\sqrt{1+\mu^2}},\end{equation}
			\begin{equation}\mathcal{M}(RR;LL)=\mathcal{M}(LL;RR)=\frac{1+\mu^2(1+\cos\theta)}{\mu^2(1+\mu^2)},\end{equation}
			\begin{equation}\mathcal{M}(_{LR}^{RL};RR)=-\mathcal{M}(_{LR}^{RL};LL)=\frac{(1+\mu^2\cos\theta)\cot(\frac{\theta}{2})}{\mu^2\sqrt{1+\mu^2}},\end{equation}
			\begin{equation}\mathcal{M}(RL;RL)=\mathcal{M}(LR;LR)=\frac{(1+\mu^2(1+\cos\theta))\cot^2(\frac{\theta}{2})}{\mu^2},\end{equation}
			\begin{equation}\mathcal{M}(RL;LR)=\mathcal{M}(LR;RL)=1-\cos\theta-\frac{1}{\mu^2}.\end{equation}
			
		}
		
		\section{Reduced density matrices}\label{B}
		From Eqs. (\ref{Eq.3.1.26})-(\ref{Eq.3.1.3}), we can compute the various reduced density matrices for the  final states
		\begin{equation}\begin{aligned}
				\rho_{AB}^{f}=\frac{1}{\mathcal Z(\theta,\mu;\eta)}\sum_{r,s,r^{\prime},s^{\prime}}&\left[\cos^2\eta\mathcal{M}(RR;rs)\mathcal{M}^{\dagger}(RR;r^{\prime}s^{\prime})|r\rangle_A|s\rangle_{BB}\langle s^{\prime}|_A\langle r^{\prime}|\right.\\
				&\left.+\sin^2\eta\mathcal{M}(RL;rs)\mathcal{M}^\dagger(RL;r^{\prime}s^{\prime})|r\rangle_A|s\rangle_{BB}\langle s^{\prime}|_A\langle r^{\prime}|\right],
			\end{aligned}
		\end{equation}
		\begin{equation}\begin{aligned}
				\rho_{AC}^{f}=\frac{1}{\mathcal Z(\theta,\mu;\eta)}\sum_{r,r^{\prime},s}&\Bigl[\mathrm{cos}^{2}\eta\mathcal{M}(RR;rs)\mathcal{M}^{\dagger}(RR;r^{\prime}s)|r\rangle_{A}|R\rangle_{CC}\langle R|_{A}\langle r^{\prime}| \\
				& +e^{-i\beta}\sin\eta\cos\eta\mathcal{M}(RR;rs)\mathcal{M}^\dagger(RL;r's)|r\rangle_A|R\rangle_{CC}\langle L|_A\langle r'| \\
				& +e^{i\beta}\sin\eta\cos\eta\mathcal{M}(RL;rs)\mathcal{M}^\dagger(RR;r^{\prime}s)|r\rangle_A|L\rangle_{CC}\langle R|_A\langle r^{\prime}| \\
				& +\sin^2\eta\mathcal{M}(RL;rs)\mathcal{M}^\dagger(RL;r^\prime s)|r\rangle_A|L\rangle_{CC}\langle L|_A\langle r^\prime|\Big],
			\end{aligned}
		\end{equation}
		\begin{equation}\begin{aligned}
				\rho_{BC}^{f}  =\frac{1}{\mathcal Z(\theta,\mu;\eta)}\sum_{r,s,s^{\prime}}&\bigl[\mathrm{cos}^{2}\eta\mathcal{M}(RR;rs)\mathcal{M}^{\dagger}(RR;rs^{\prime})|s\rangle_{B}|R\rangle_{CC}\langle R|_{B}\langle s^{\prime}| \\
				& +e^{-i\beta}\sin\eta\cos\eta\mathcal{M}(RR;rs)\mathcal{M}^\dagger(RL;rs^\prime)|s\rangle_B|R\rangle_{CC}\langle L|_B\langle s^\prime| \\
				& +e^{i\beta}\sin\eta\cos\eta\mathcal{M}(RL;rs)\mathcal{M}^\dagger(RR;rs^{\prime})|s\rangle_B|L\rangle_{CC}\langle R|_B\langle s^{\prime}| \\
				& +\sin^2\eta\mathcal{M}(RL;rs)\mathcal{M}^\dagger(RL;rs^\prime)|s\rangle_B|L\rangle_{CC}\langle L|_B\langle s^\prime|\Big].
			\end{aligned}
		\end{equation}
		The single-particle reduced density matrices $\rho_A^{f},\rho_B^{f},\rho_C^{f}$ can be obtained by taking one further partial trace.

	\end{widetext}

	\newpage
	\appendix 
\end{document}